\documentclass[iop]{emulateapj}

\usepackage{tabularx}

\newcommand{\myemaila}{satoko.ss@nao.ac.jp}
\newcommand{\myemailb}{sss@mx.ibaraki.ac.jp}

\slugcomment{To appear in ApJ Letters}

\shorttitle{HCN Absorption Features in NGC 1052}
\shortauthors{Sawada-Satoh et al.}

\begin{document}


\title{Spatially Resolved HCN Absorption Features \\ 
in the Circumnuclear Region of NGC 1052}


 \author{%
   Satoko Sawada-Satoh\altaffilmark{1,2}, 
   Duk-Gyoo Roh\altaffilmark{3}, 
   Se-Jin Oh\altaffilmark{3}, 
   Sang-Sung Lee\altaffilmark{3,4}, 
   Do-Young Byun\altaffilmark{3,4}, \\
   Seiji Kameno\altaffilmark{5,6},
   Jae-Hwan Yeom\altaffilmark{3},
   Dong-Kyu Jung\altaffilmark{3},
  Hyo-Ryoung Kim\altaffilmark{3},
  Ju-Yeon Hwang\altaffilmark{3,7} 
    }  

\altaffiltext{1}{Mizusawa VLBI Observatory, National Astronomical Observatory of Japan,
2-12 Hoshigaoka-cho, Mizusawa-ku, Oshu, Iwate 023-0861, Japan; \myemaila}
\altaffiltext{2}{Center for Astronomy, Ibaraki University, 2-1-1 Bunkyo, Mito, Ibaraki 310-8512, Japan; \myemailb}
\altaffiltext{3}{Korea Astronomy and Space Science Institute, 
776 Daedeok-daero, Yuseong, Daejeon 34055, Republic of Korea}
\altaffiltext{4}{University of Science and Technology, 
217 Gajeong-ro, Yuseong-gu, Daejeon 34113, Republic of Korea }
\altaffiltext{5}{Joint ALMA Observatory, Alonso de Cordova 3107 Vitacura, Santiago 763 0355, Chile}
\altaffiltext{6}{National Astronomical Observatory of Japan, 2-21-1 Osawa, Mitaka, Tokyo 181-8588, Japan}
\altaffiltext{7}{SET system, 16-3 Gangnam-daero 8-gil, Seocho-gu, Seoul 06787, Korea}


\begin{abstract}
We present the first VLBI detection of HCN molecular absorption 
in the nearby active galactic nucleus NGC~1052. 
Utilizing the 1 mas resolution achieved by the Korean VLBI Network, 
we have spatially resolved the HCN absorption 
against a double-sided nuclear jet structure. 
Two velocity features of HCN absorption are detected significantly 
at the radial velocity of 1656 and 1719 km~s$^{-1}$, 
redshifted by 149 and 212 km~s$^{-1}$ with respect to 
the systemic velocity of the galaxy. 
The column density of the HCN molecule is estimated to be 
$10^{15}$---$10^{16}$ cm$^{-2}$, 
assuming an excitation temperature of 100---230 K. 
The absorption features show high optical depth localized on the receding jet side, 
where the free-free absorption occurred due to the circumnuclear torus. 
The size of the foreground absorbing molecular gas is estimated to be on
approximately one-parsec scales, 
which agrees well with the approximate size of the circumnuclear torus. 
HCN absorbing gas is likely to be several clumps smaller than 0.1 pc 
inside the circumnuclear torus. 
The redshifted velocities of the HCN absorption features imply 
that HCN absorbing gas traces ongoing infall motion inside the circumnuclear torus 
onto the central engine.

\end{abstract}





\section{Introduction}

According to the most common active galactic nuclei (AGNs) model,
i.e. the Unified Scheme of AGNs \citep{antonucci93}, 
AGNs consist of a supermassive black hole (SMBH) or a central engine with a 
rotating dense gas disk or torus surrounding the SMBH. 
It is generally accepted that 
pronounced activity in an AGN 
is generated by accreting gas from its torus onto a SMBH. 
The distribution and kinematics of the 
circumnuclear gas are keys for understanding the fueling of the AGN. 

High angular resolution studies of molecular gas 
in the center of the external galaxies  (radius of $\sim$ 1 kpc) 
have been obtained with millimeter interferometers.  
The size of the circumnuclear torus, however, 
is smaller than 10 pc \citep[e.g.][]{garcia16}, 
and a milliarcsecond (mas) resolution is required 
to study its internal structure in nearby AGNs. 
While conventional millimeter interferometers, even ALMA, did not achieve
such a high angular resolution, 
VLBI observations have revealed the parsec- or subparsec-scale morphology 
of nearby AGNs. 
Generally, thermal emission lines from molecular gas are not luminous 
enough to detect with the VLBI.
VLBI maps, however, can display thermal absorption lines of the gas in silhouette
against a bright background synchrotron radiation source 
with a mas resolution.

NGC 1052 is a nearby elliptical galaxy with a systemic velocity 
($V_{\rm sys} = cz$) of 1507 km~s$^{-1}$ \citep{jensen03},  
implying a distance of 20~Mpc assuming $H_0= 75$ km~s$^{-1}$~Mpc$^{-1}$
and $q_0 = 0.5$. 
Its nuclear activity is classified as occurring within a low-ionization 
nuclear emission-line region \citep[LINER; e.g.][]{gabel00}.
NGC 1052 has a nearly symmetric double-sided radio jet structure 
along the east--west direction 
from kiloparsec to parsec scales
\citep[e.g.][]{jones84, wrobel84, kellermann98}. 
Past VLBI studies constrained the jet inclination angle 
to lie $\geq 57^{\circ}$
using the optically thin radio flux density of the twin jet \citep{kadler04, sss08, baczko16}. 
The parsec-scale jet structure shows a prominent gap 
between 
the eastern (approaching) and western (receding) jets 
at various centimeter wavelengths 
\citep[e.g.][]{claussen98, vermeulen03}. 
VLBI images at 43~GHz, however, show 
an innermost component in the gap 
\citep{kadler04, sss08}. 
\cite{kameno01} proposed the existence of a parsec-scale 
circumnuclear torus consisting of cold dense plasma 
from the convex radio continuum spectra 
due to the free-free absorption (FFA) by the foreground plasma. 

Besides the ionized gas, 
several atomic and molecular gases are also found 
toward the center of NGC 1052. 
An H$_2$O megamaser emission is detected at velocities 
by redshifted 50---350 km~s$^{-1}$ 
with respect to $V_{\rm sys}$ 
\citep{braatz94, braatz03, kameno05}, 
and the H$_2$O maser gas lies on the inner components 
in the center of NGC~1052, where the plasma torus is obscured 
\citep{claussen98, sss08}. 
Neutral atomic hydrogen (\ion{H}{1}), OH, HCO$^+$, HCN, CO 
transitions are found toward the center of NGC~1052 
as absorptions 
\citep{vangorkom86, omar02, liszt04, impellizzeri08evn}. 
The velocity range over which those absorption lines are observed 
spans from 1500 to 1800 km~s$^{-1}$, 
redshifted with respect to $V_{\rm sys}$-like H$_2$O maser emission. 
These facts imply that 
ionized, neutral, and molecular gases coexist  
in the vicinity of the central engine.

To investigate the geometry and physical properties of the molecular gas 
in the circumnuclear region of NGC~1052, 
we have performed high-resolution observations toward 
HCN (1--0) absorption lines 
with the Korean VLBI Network (KVN). 
Here we show the first parsec-scale HCN (1--0) maps of the center of NGC~1052. 
One milliarcsecond corresponds to 0.095 pc in the galaxy.

\section{Observations and data reduction}

KVN observations of NGC~1052 were carried out in 2015 March, 
for a total on-source time of 7.5 hr. 
To achieve high sensitivity for the 3~mm HCN (1--0) absorption line observation, 
simultaneous dual-frequency data 
at the K (21700---21828 MHz) and W (88643---88871 MHz) bands 
were recorded 
using the KVN multi-frequency receiving system \citep{oh11, han13}. 
Two IFs of left hand circular polarization with a bandwidth of 128 MHz were used 
simultaneously. 
One was assigned to the W-band for HCN (1--0), 
and the other was tuned to the K-band, the reference frequency 
to track a rapid phase time variation at the W-band. 
The velocity coverage of one IF at the W-band was $\sim$ 400 km~s$^{-1}$. 
NRAO~150 was also observed as a calibrator for phase and bandpass.
The Mark5B system at a recording rate of 1024 Mbps was used for recording the data.
The correlation was processed at the Korea--Japan Correlation Center \citep{yeom09, lee15a}. 
We corrected the visibility amplitude decrement 
due to the digital quantization loss and the characteristics of the KJCC, 
using the method of \cite{lee15b}.

Calibration, data flagging, fringe fitting, and imaging were performed using the NRAO AIPS software.
To track a rapid phase time variation at the W-band,  
we analyzed these data using the frequency phase transfer method, 
in which the phase solutions of the low-frequency band (K-band) are transferred 
to the high-frequency (W-band) 
by scaling by their frequency ratios  
\citep{middelberg05, rioja11}. 
Doppler correction was applied using the AIPS task CVEL. 
The continuum was subtracted with the AIPS task UVLSF 
in the ($u, v$) domain. 
We corrected the visibility phase 
using self-calibration with the averaged line-free channels, 
and applied the result of the self-calibration 
in the absorption line images.
Therefore, the absorption and continuum images could be accurately overlaid 
with positional errors only limited by the thermal noise in the absorption images. 
Images were produced with natural weighting, and 
the resulting FWHM of the synthesized beam is $1.44\times0.90$ mas
($0.137\times0.085$ pc in NGC~1052).

\section{Results} \label{sec:results}

\begin{figure}[th]
\epsscale{1.0}
\plotone{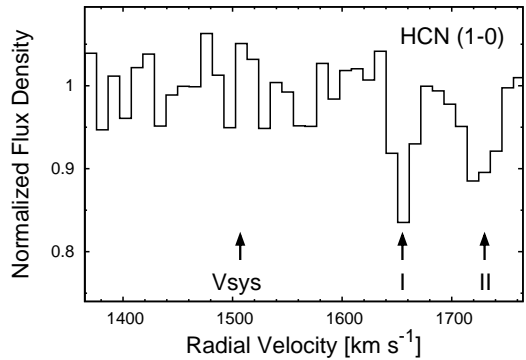}
\caption{Spectral profile of the HCN (1--0) absorption line 
toward the center of NGC~1052 obtained with the KVN. 
Two velocity features of HCN absorption were detected 
at 1656 and 1719 km~s$^{-1}$. 
We gave labels to the former (I) and the latter (II), respectively. 
The rms noise in the normalized flux density is 0.035. 
\label{spc}}
\end{figure}

\begin{figure}[th]
\epsscale{1.0}
\plotone{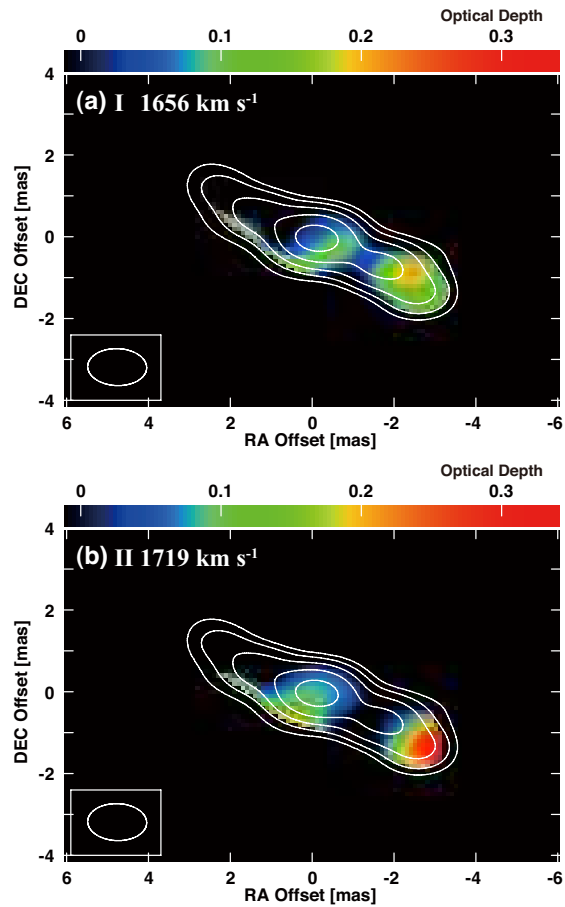}
\caption{Color images of the HCN (1--0) optical depth of (a) feature I 
(1656 km~s$^{-1}$) 
and (b) feature II (1719 km~s$^{-1}$) 
in the circumnuclear region of NGC~1052, 
overlaid by a contour map of 89 GHz continuum emission.  
The contour starts at the $3 \sigma$ level, increasing by a factor of 2, 
where $\sigma=$ 3.9 mJy~beam$^{-1}$. 
The absorption maps have achieved 
a 1 $\sigma$ noise level of 4.6 mJy beam$^{-1}$ in intensity, 
or $\sim$0.02 in optical depth. 
The velocity width of one channel map is 52.9 km~s$^{-1}$.
\label{hcnmap}}
\end{figure}

The HCN (1--0) absorption spectra 
toward NGC~1052 obtained with the KVN are shown in Figure~\ref{spc}. 
The velocity resolution is 10.5 km~s$^{-1}$. 
Two velocity features of HCN absorption are detected, labeled as I and II. 
The deepest absorption feature, I, shows a peak at 1656 km~s$^{-1}$ channel 
and reaches $-16 \%$ of the continuum level. 
The second-deepest absorption feature, II, reaches $-10\%$ of the continuum level
 at 1719 km~s$^{-1}$.  
The two absorption features are also found  
in the absorption profile from the PdBI data \citep{liszt04}. 
The velocity widths and the peak velocities agree well with 
the profile measured using the PdBI, 
but the absorption depths in the spectral profile with the KVN 
are deeper than those of the PdBI (2---6$\%$). 
Although \Citet{liszt04} have shown several other absorption features 
with the depth of $\sim-3\%$ 
around 1500---1600 km~s$^{-1}$ 
from the PdBI data, 
our KVN data  did not detect them significantly.

Figure~\ref{hcnmap} is a superposition of 
the line-free continuum image at 89 GHz 
and the opacity image for each HCN absorption feature. 
The continuum image reveals that 
a nearly symmetric double-sided radio jet structure 
along the east--west direction is elongated over $\sim$0.5 pc 
(contour map in Figure~\ref{hcnmap}). 
The brightest continuum component is seen at the center of the elongated structure. 
The sum of all CLEAN components within 1.4 mas 
(the major axis of the synthesized beam) 
from the peak is 332 mJy,
and it is $\sim 80 \%$ of the value of the central continuum component at 86 GHz 
measured in 2004 October  
by the Global millimeter-VLBI Array \citep{baczko16}. 
The flux densities for the east and west jet are estimated to be 
49 and 180 mJy, respectively. 
HCN opacity distribution against the double-sided radio jet structure shows that 
high opacities ($>0.2$) are clearly found on the western receding jet side. 
Each opacity peak position of the absorption features I and II  
is offset by 0.24 and 0.27 pc, respectively, from the continuum peak position,
where the central engine is supposed to exist.

\section{Discussions}

\begin{table*}[t]
\begin{center}
\caption{Column density of HCN (1--0) absorption.\label{tb:hcn}}
\begin{tabular}{lcccccc}
\tableline
\tableline
Label & $V_{\rm p}$ & $V_{\rm p} - V_{\rm sys}$ & $\Delta v$ & 
$\int \tau dv$ & $N_{\rm HCN,100}$ & $N_{\rm HCN,230}$ \\
   & [km~s$^{-1}$] & [km~s$^{-1}$] & [km~s$^{-1}$] & [km~s$^{-1}$] & [10$^{14}$ cm$^{-2}$] & [10$^{14}$ cm$^{-2}$] \\
  (1) & (2) & (3) & (4) & (5) & (6) & (7) \\
\tableline
I & 1656 & 149 & 31.7 & $9.3\pm0.6$ & $9.5\pm0.6$ & $50\pm3$ \\
II & 1719 & 212 & 52.9 & $19\pm1$ & $20\pm1$ & $101\pm5$ \\
\tableline
\end{tabular}
\tablecomments{(1) Absorption feature ID, shown in Figure~\ref{spc}. 
(2) Peak velocity of the absorption feature.
(3) Velocity with respect to $V_{\rm sys}$.
(4) Line width. 
(5) Velocity-integrated optical depth. 
(6) HCN column density with $T_{\rm ex}=100$ K.
(7) HCN column density with $T_{\rm ex}=230$ K.}
\end{center}
\end{table*}

\subsection{Absorbing Molecular Gas Properties}

The HCN opacity obtained from our observations was beyond 0.1
in the circumnuclear region of NGC~1052, 
and one order higher than that from the PdBI observations \citep{liszt04}. 
This indicates that the HCN covering factor is 
much smaller on the scale of a few hundreds parsecs.
We estimated the mean opacity 
over the whole parsec-scale source structure $\left< \tau \right>$, 
using 
\begin{equation}
\left< \tau \right> = 
\frac{\displaystyle \int \!\! \int \tau(x,y) I(x,y) \,dx \,dy}{\displaystyle \int \!\! \int I(x,y) \,dx \,dy}
\end{equation}
where $\tau(x,y)$ is the HCN opacity distribution shown in the color scale of Figure~\ref{hcnmap}, 
and $I(x,y)$ is the 89 GHz continuum image. 
The estimated $\left< \tau \right>$ is 0.027 and 0.028 
for the absorption features I and II, respectively. 
They are on the same order of HCN opacity yielded from the normalized flux of 0.98 and 0.94  
for the absorption features I and II observed with the PdBI \citep{liszt04}. 
Thus, the HCN covering factor is likely to be $\sim1$ on parsec scales.

Under the assumption of the local thermodynamic equilibrium, 
and a covering factor of 1, 
the total molecular column density of the J=1--0 absorption line 
is derived as 
\begin{equation}
N_{\rm tot} = 
\frac{3 k}{8 \pi^3 \mu^2 B} ~ 
\frac{(T_{\rm ex} + hB /3k)}{\bigl[1-\exp{\bigl( - h\nu/ kT_{\rm ex}\bigr) \bigr]}} 
\int \tau dv 
\label{eq:nhcn}
\end{equation}
where  
$k$ is the Boltzmann constant, 
$h$ is the Planck constant, 
$\mu$ is the permanent dipole moment of the molecule, 
$B$ is the rotational constant, 
$T_{\rm ex}$ is the excitation temperature, and 
$\int \tau dv$ is the velocity-integrated optical depth of the absorption feature.

Using the equation~(\ref{eq:nhcn}),  
$\mu=2.98$ Debye, $B= 44315$ MHz for HCN, 
and assuming $\int \tau dv = \tau_{\rm max} \Delta v$,  
where $\tau_{\rm max}$ is the maximum of optical depth, 
we give the total column density of HCN 
for each absorption feature in Table~\ref{tb:hcn}. 
To determine the column density of molecular hydrogen (H$_2$), 
we assume a HCN-to-H$_2$ abundance ratio of 10$^{-9}$, 
from the HCN-to-H$_2$ abundance ratio estimates of (0.2---4.1)$\times10^{-9}$ 
of several HCN clumps in the Galactic Circumnuclear Disk \citep[CND;][]{smith14}. 
Here we take $T_{\rm ex}=100$ K and 230 K, 
as recent millimeter and submillimeter interferometric observations 
have detected the vibrationally excited HCN emission lines 
in the dust-enshrouded AGN of the luminous infrared galaxies 
\citep{sakamoto10, imanishi13}.  
The derived H$_2$ column density ($N_{\rm H_2}$) 
ranges $10^{24}$---$10^{25}$  cm$^{-2}$, 
using the values of $N_{\rm HCN}$ in Table~\ref{tb:hcn}. 
Hence, equivalent $N_{\rm H}$ would be of the same order as $N_{\rm H_2}$. 
It is one or two orders higher than 
 the estimate of the $N_{\rm H} \sim 10^{23}$ cm$^{-2}$ 
from modeling of X-ray spectra \citep{guainazzi99, weaver99, guainazzi00, brenneman09}, 
and the electron column density of 
$\sim10^{23}$  cm$^{-2}$ 
estimated from the FFA opacity of the dense plasma \citep{sss09}.

\subsection{Circumnuclear torus model}

\begin{figure*}[th]
\epsscale{0.9}
\plotone{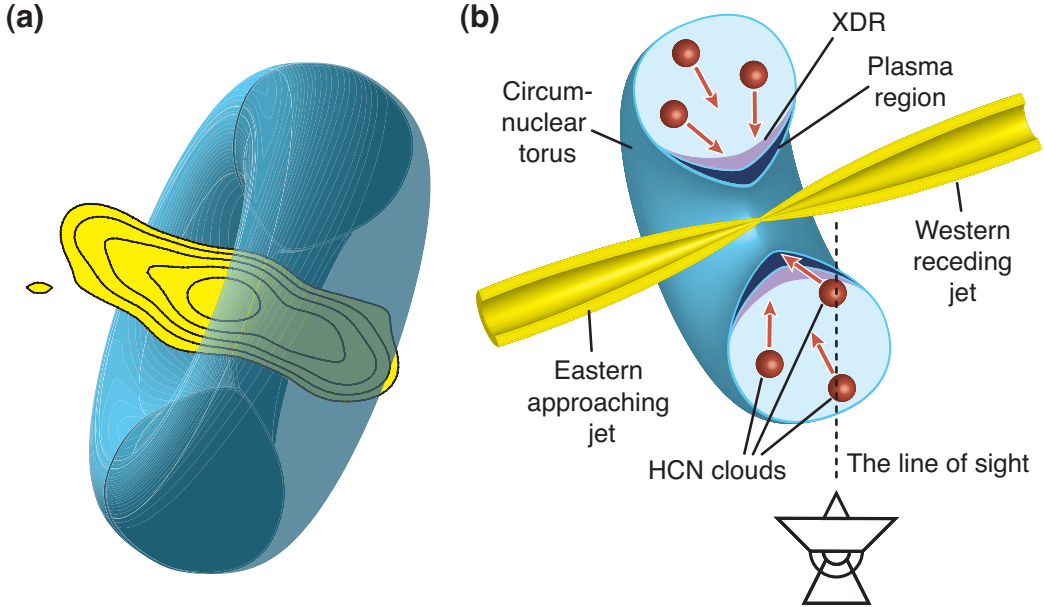}
\caption{(a) Possible model of the oriented double-sided jet and 
the circumnuclear torus. 
The near side of the torus covers the receding jet side. 
(b) Schematic diagram for the intersection of the circumnuclear torus 
in NGC~1052. 
The torus has two-phase layers on the inner surface, 
the X-ray heated plasma region and the X-ray dissociation region (XDR). 
HCN absorbing gas is located in the cooler molecule region next to XDR. 
HCN absorbing gas could be clumpy, 
and the line of sight passes through several clumpy clouds 
with a different radial velocity. 
\label{model}}
\end{figure*}

\cite{kameno05} and \cite{sss08} proposed that 
the circumnuclear torus consists of several phase layers. 
On the inner surface of the torus, 
a hot ($\sim8000$ K) plasma layer is formed by X-ray emission 
from the central engine and 
it causes FFA. 
A heated (above $\sim400$ K) molecular layer or X-ray dissociation region 
\citep[XDR;][]{neufeld94, neufeld95} lies next to the plasma layer, 
and the H$_2$O megamaser emission arises from here. 
As well as FFA opacity distribution, 
the HCN opacity shows high values on the western receding side of the jet. 
Thus, HCN absorbing gas could be associated with the torus, 
like the dense plasma. 
If $T_{\rm ex}$ of HCN is $\sim230$ K, 
HCN molecules should lie in the cooler molecule layer next to XDR. 
The redshifted HCN absorption against the continuum emission 
is likely indicative of ongoing gas infall. 
The velocities of the HCN absorption features are close to 
the centroid velocity of the broad H$_2$O maser emission ($\sim1700$ km~s$^{-1}$). 
Therefore, H$_2$O and HCN could trace the same infall motion inside the torus. 
The HCN absorption spectrum consists of at least two narrow absorption features, and 
HCN absorbing gas is more likely to be several small ($\le$ 0.1 pc) clumps or layers with a different velocity, 
rather than a large homogeneous structure with a single velocity.
We have seen the opacity contrast between the absorption features I and II, and 
the contrast also suggests the inhomogeneity inside the molecular torus. 
In Figure~\ref{model}, we present a possible model for the geometry of the circumnuclear torus 
and the jet in NGC 1052.  
Since the jet axis is inclined from the sky plane, 
the near side of the torus should cover the receding jet side. 
The line of sight intersects at least two HCN gas clumps inside the torus. 

As we have mentioned in Section \ref{sec:results}, 
there is no significant detection of absorption features around 1500---1600 km~s$^{-1}$, 
close to $V_{\rm sys}$. 
A possible explanation is that 
the missing absorption features could arise 
not from the compact molecular clumps in the circumnuclear region, 
but from the foreground diffuse interstellar medium in the host galaxy. 
If so, their covering factor would not vary on scales 
between a few parsecs and a few hundreds parsecs, 
and thus the depth of the absorption features would be $\sim -3\%$, 
the same as the PdBI results.  
It is comparable to the rms noise level of our spectral profile with the KVN, 
and no significant detection comes as a consequence.
Furthermore, 
when their background continuum sources have a faint and extended structure, 
the background continuum emissions should be fully resolved-out. 
Therefore, 
the absorption features against the resolved-out background sources would be invisible in our KVN data.

We can estimate an approximate size of the foreground absorbing molecular gas 
using the relation $N_{\rm H_2} = f_v n_{\rm H_2} L $, 
where 
$f_v$ is the volume filling factor, 
$n_{\rm H_2}$ is the molecular hydrogen volume density,  
and 
$L$ is the size of the molecular gas. 
Here we simply assume $f_v=$ 0.01. 
Adopting $N_{\rm H_2}$ of $10^{24}$--$10^{25}$ cm$^{-2}$ from our results and 
$n_{\rm H_2}$=(0.1---2)$\times10^6$ cm$^{-3}$ in the Galactic CND \citep{smith14}, 
the resulting $L$ is approximately on 1-pc scales. 
It is consistent with the approximate size of the circumnuclear torus of NGC~1052 \citep{kameno01}. 
On the other hand,  
the size of the innermost receding jet component can be estimated 
as 0.2---0.4 mas from the VLBA images at 43 GHz \citep{kadler04, sss08}, and 
it corresponds to 0.02---0.04 pc in linear scale.

Based on our measurements of 
the radial velocities and the distribution of HCN absorption features, 
the mass infall rate of the gas accretion 
toward the central engine can be estimated by 
$\dot{M} = f_v R_{\rm in}^2  \rho V_{\rm in} \Omega$, 
where
$f_v$ is the volume filling factor,  
$R_{\rm in}$ is an infall radius, 
$\rho$ is the mass density of the infalling materials, 
$V_{\rm in}$ is the infall velocity, 
and $\Omega$ is the solid angle of the torus from the center.
Here we assume 
$\rho = N_{\rm H} m_{\rm H} / R_{\rm in}$, 
where 
$N_{\rm H}$ is the column density of a hydrogen atom and 
$m_{\rm H}$ is the mass of a hydrogen atom. 
Giving $R_{\rm in}=$ 1 pc, 
$N_{\rm H}=$ $10^{24}$---$10^{25}$ cm$^{-2}$,
and 
$V_{\rm in}=$ 200 km~s$^{-1}$ 
as the approximate velocity of the HCN absorption feature 
with respect to $V_{\rm sys}$ 
($V_{\rm p}-V_{\rm sys}$ in Table~\ref{tb:hcn}), 
the derived mass infall rate ranges 
$\dot{M} =$ (47---470)  $f_v (\Omega / 4\pi) M_{\sun}$ yr$^{-1}$. 
If ($\Omega/ 4 \pi$) takes a few 0.1 and $f_v$ is 0.01, 
it would be comparable to the calculation of the accretion rate 
of $10^{-1.39} M_{\sun}$ yr$^{-1}$ 
from the hard X-ray luminosities by \cite{wu06}.
We note that the X-ray luminosity indicates an instantaneous accretion rate. 
However, our estimation of the accretion rate in the molecular torus gives 
a long-term ($R_{\rm in}/V_{\rm in} \sim 5000$ year) activity. 
The coincidence of the two accretion rate values suggests 
the continuity of AGN activity in NGC~1052.

\section{Summary}

We have conducted 1 mas angular-resolution observations 
toward the HCN(1--0) absorption of the circumnuclar region 
in NGC~1052 with the KVN. 
Two HCN absorption features are identified and 
reach a depth of $\ge 10\%$ from the continuum level 
at 1656 and 1719 km~s$^{-1}$, 
redshifted with respect to $V_{\rm sys}$. 
We find an $N_{\rm HCN}$ of $10^{15}$---$10^{16}$ cm$^{-2}$ in the center of NGC~1052, 
assuming an HCN covering factor of 1 and a $T_{\rm ex}$ of 100---230 K. 
The HCN opacity distribution against the double-sided radio continuum jet structure 
shows a localization of the HCN high opacity on the receding jet side. 
The size of the foreground absorbing molecular gas is estimated to be 
approximately on 1-pc scales. 
We argue that these results are most naturally explained 
by a circumnuclear torus that consist of several phase layers,
perpendicular to the jet axis of NGC~1052. 
HCN absorbing gas distribution could be clumpy,
and  could be 
associated with the cooler molecular region 
inside the torus. 
The redshifted velocities of the HCN absorption features 
account for the infall motion onto the central engine, 
such as an H$_2$O megamaser emission.



\acknowledgments

We acknowledge the anonymous referee who provided constructive suggestions. 
We are grateful to all members of  KVN for supporting our observations. 
S.S.S. would like to thank Yoshiharu Asaki and Taehyun Jung 
for their helpful discussions. 
S.S.L. was supported by the National Research Foundation of Korea(NRF) grant funded 
by the Korea government(MSIP) (No. NRF-2016R1C1B2006697).
The KVN is a facility operated by the Korea Astronomy and Space Science Institute. 
The KVN operations are supported 
by the Korea Research Environment Open NETwork,  
which is managed and operated by Korea Institute of Science and Technology Information.



{\it Facilities:} \facility{KVN}.

\end{document}